\documentclass[10pt,showpacs,twocolumn]{revtex4}
\usepackage{graphicx}
\usepackage{amsmath}
\usepackage{amssymb}
\usepackage{dsfont}

\begin{document}
\title{Molecular orbital tomography beyond the plane wave approximation}
\author{Yang Li,$^{1,2}$ Xiaosong Zhu,$^{1,2}$ Pengfei Lan,$^{1,2}$\footnote{pengfeilan@hust.edu.cn} Qingbin Zhang,$^{1,2}$ Meiyan Qin,$^{1,2}$ and Peixiang Lu$^{1,2}$\footnote{lupeixiang@hust.edu.cn}}
\affiliation{$^{1}$School of Physics and Wuhan National Laboratory for Optoelectronics, Huazhong University of Science and Technology, Wuhan, 430074, China\\
$^{2}$Key Laboratory of Fundamental Physical Quantities Measurement of Ministry of Education, Huazhong University of Science and Technology, Wuhan 430074, China
}
\date{\today}

\begin{abstract}
The use of plane wave approximation in molecular orbital tomography via high-order harmonic generation has been questioned since it was proposed, owing to the fact that it ignores the essential property of the continuum wave function. To address this problem, we develop a theory to retrieve the valence molecular orbital directly utilizing molecular continuum wave function which takes into account the influence of the parent ion field on the continuum electrons. By transforming this wave function into momentum space, we show that the mapping from the relevant molecular orbital to the high-order harmonic spectra is still invertible. As an example, the highest orbital of $\mathrm{N_2}$ is successfully reconstructed and it shows good agreement with the \emph{ab initio} orbital. Our work clarifies the long-standing controversy and strengthens the theoretical basis of molecular orbital tomography.

\end{abstract}
 \maketitle

The fast development of strong-field physics has provided versatile perspectives for probing the structure and ultrafast electron dynamics in atoms and molecules with attosecond and {\AA}ngst\"{o}rm resolutions \cite{Pfeiffer,Lein,Smirnova,Wu,Vozzi1}. A fascinating application, known as molecular orbital tomography (MOT) based on high-order harmonic generation (HHG), has attracted a great deal of attention for its potential use of observing chemical reactions in molecules by directly imaging the valence molecular orbital \cite{Itatani,Haessler1,Haessler2,Salieres,Vozzi2,Zwan,Qin,Zhu2,Chen}. Following the pioneering work by Itatani \emph{et al.} \cite{Itatani} which successfully reconstructed the highest occupied molecular orbital (HOMO) of $\mathrm{N_2}$ using high-order harmonic spectra from aligned molecules, MOT has been extended to more complex species such as $\mathrm{CO_2}$ \cite{Vozzi2} and asymmetric molecules of $\mathrm{HeH^{2+}}$ \cite{Zwan} and $\mathrm{CO}$ \cite{Qin,Chen}.

The original MOT theory is based on the plane wave approximation (PWA), which assumes that the continuum wave functions are unperturbed by the electron interaction with the parent ion and can be viewed as plane waves \cite{Itatani}. With this assumption, the transition dipole is given in the form of the Fourier transform of the HOMO weighted by the dipole operator. Thus by performing inverse Fourier transform, the HOMO of the molecule can be reconstructed. However, it is a drastic simplification to represent continuum wave functions of a molecule by plane waves, especially in the low-energy region where most HHG experiments are performed. Many effects during the rescattering process in HHG are not treated properly such as the distortion of the continuum wave function due to the molecular potential. Recent works demonstrated that HHG from molecules was influenced by the Coulomb potential of the parent ion and some features of high-order harmonics are attributable to the distortion of continuum wave function \cite{Zwan2,Paul,Zhu3}. Thereby, the tomographical image of the molecular orbital can be greatly modulated. All these features make the foundation of original MOT procedure unstable. For this reason, the theoretical foundation of MOT has been questioned \cite{Lin1,Lin2,Walters} since it was proposed. Therefore, a method to correct these deviations is highly desirable.

In this paper, we revisit the MOT and develop a tomographic theory in which molecular continuum wave functions can be directly used to retrieve the valence molecular orbital. By using the momentum-space representation of the continuum waves, we show that the mapping from the relevant molecular orbital to the high-order harmonic spectra is still invertible. As an example, we reconstruct the symmetric $3\sigma_{g}$ HOMO of $\mathrm{N_2}$ molecule by using two-center Coulomb waves (TCC) as the continuum wave function within this theory. The results show that the main features of the HOMO are quite well reproduced and quantitative agreement between the retrieved orbital and the \emph{ab intio} one is achieved.

The MOT procedure \cite{Itatani} is performed by firstly aligning the molecules using a laser pulse and then focusing a second, more intense, pulse on the aligned molecules to generate high-order harmonics. By changing the relative angle between the molecular frame and the polarization vector of the laser pulse, harmonic spectra are obtained at different orientations of the molecules. The high-order harmonic emission rate with harmonic frequency $\omega$ is given by
\begin{eqnarray}
\Gamma(\omega)\varpropto\omega^4|a(\omega)\mathbf{d}(\mathbf{k})|^2,
\end{eqnarray}
with $\mathbf{d}(\mathbf{k})=\langle \psi_{0}(\mathbf{r})|\mathbf{r}|\psi_{\mathbf{k}}(\mathbf{r})\rangle$ being the transition dipole matrix element in momentum space between a continuum wave function $\psi_{\mathbf{k}}$ and the valence orbital $\psi_{0}$ of the target molecule. The complex amplitude of the continuum state $a(\omega)$ can be obtained by recording the spectrum from a reference atom with the same ionization energy as the target molecule and dividing by the calculated transition dipole matrix element for the ground state of the atom. Once $a(\omega)$ is factored out, The modulus of the transition matrix elements $\mathbf{d}(\mathbf{k})$ can be obtained according to Eq. (1). The dipole phases can be recovered by perform a series of RABBIT measurements of the HHG emission with a set of alignment angles \cite{Haessler1}.

At its heart, the tomographic algorithm relies on the obtaining of the transition matrix elements $\mathbf{d}(\mathbf{k})$. It can be written in the length form as
\begin{eqnarray}
\mathbf{d}(\mathbf{k})=\int d^{3}\mathbf{r} \psi_{\mathbf{k}}(\mathbf{r}) \mathbf{r} \psi_{0}(\mathbf{r}).
\end{eqnarray}
In the original tomographic procedure, the use of PWA is essential. The continuum wave function is given by plane waves
\begin{eqnarray}
\psi_{\mathbf{k}}^{PW}(\mathbf{r})=(2\pi)^{-3/2}\exp(i\mathbf{k}\cdot\mathbf{r}).
\end{eqnarray}
With this approximation, the ground state wave function could be reconstructed by performing inverse Fourier transform
\begin{eqnarray}
\mathbf{r} \psi_{0}(\mathbf{r})=(2\pi)^{-3/2}\int d^{3}\mathbf{k} \mathbf{d}(\mathbf{k}) \exp(-i\mathbf{k}\cdot\mathbf{r}).
\end{eqnarray}
\begin{figure}
\centerline{
\includegraphics[width=8cm]{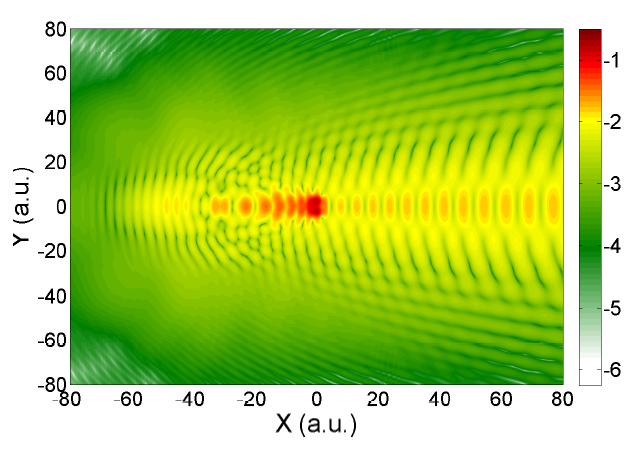}}
\caption{Two-dimensional time-dependent Schr\"{o}dinger equation simulations of the re-collision electron wave packet with a $\mathrm{H_{2}^{2+}}$ molecular ion at the instance when the wave packet approaches the nuclei from the right-hand side. The $\mathrm{H_{2}^{2+}}$ molecular ion is aligned along the y direction The internuclear distance is $1.052 \mathrm{{\AA}}$. The
wave packet is generated by ionization of $\mathrm{H_{2}^{+}}$  with a sine laser pulse perpendicular to the molecular axis. The wavelength of the laser pulse is $1200$ $\textrm{nm}$ and the intensity is $1.0\times 10^{14}\mathrm{W/cm^2}$. }
\end{figure}
The plane wave representation of the eigenfunctions of the re-collision electron is a drastic approximation, which neglects the influence of the effective local potential. It is not adequate for the description of electron scattering states at low energies. In this energy region (from $~20$ $\textrm{eV}$ to $~ 1$ $\textrm{keV}$) where most HHG experiment is performed, the Coulomb potential of the parent ion experienced by the electron is comparable to the scattering energy, thus the measured dipole will deviate from the Fourier transform of the molecular orbital weighted by the dipole operator. Figure 1 presents the two-dimensional time-dependent Schr\"{o}dinger equation simulations of the re-colliding electron wave packet at the instance when the continuum wave function returns to the parent ion. As one can see, the re-colliding wave packet shows clear distortions from the plane waves as the electron approaches the nuclei. Part of the wave packet is distorted under the influence of the Coulomb potential and then presents a typical scattering wave character. In this case, the Fourier transform relation between $\mathbf{r} \psi_{0}(\mathbf{r})$ and $\mathbf{d}(\mathbf{k})$ in the initial theoretical formulations of MOT is broken. As a result, a more reliable tomographic theory beyond the PWA is highly desirable.

Below, we will develop a theory to retrieve the molecular orbital directly using continuum wave functions. Let us begin by defining
\begin{eqnarray}
\varphi_{\mathbf{k}}(\mathbf{k'})=(2\pi)^{-3/2}\int d^{3}\mathbf{r} \psi_{\mathbf{k}} (\mathbf{r}) \exp(-i\mathbf{k'}\cdot\mathbf{r}),
\end{eqnarray}
which satisfies
\begin{eqnarray}
\psi_{\mathbf{k}}(\mathbf{r})=(2\pi)^{-3/2}\int d^{3}\mathbf{k'} \varphi_{\mathbf{k}} (\mathbf{k'}) \exp(i\mathbf{k'}\cdot\mathbf{r}),
\end{eqnarray}
where $\varphi_{\mathbf{k}}(\mathbf{k'})$ is the momentum-space representation of the continuum wave function. Inserting Eq. (6) into Eq. (2), we can obtain
\begin{eqnarray}
\mathbf{d}(\mathbf{k})&=&(2\pi)^{-3/2} \int d^{3}\mathbf{k'}\varphi_{\mathbf{k}}(\mathbf{k'})\nonumber\\&&\times[\int d^{3}\mathbf{r}  \psi_{0}(\mathbf{r})\mathbf{r}\exp(i\mathbf{k'}\cdot\mathbf{r})],
\end{eqnarray}
where $\int d^{3}\mathbf{r}  \psi_{0}(\mathbf{r})\mathbf{r}\exp(i\mathbf{k}\mathbf{r})$ is the quantity that we need to reconstruct the molecular orbital.

In Eq. (7), $\varphi_{\mathbf{k}}(\mathbf{k'})$ defines a Fourier-space mapping from the desired quantity to the transition dipole. Generally, this mapping is not diagonal for the reason that the continuum states will have the non-zero component at $\mathrm{k'}\neq\mathrm{k}$ due to the distortions by the molecular potential. So this mapping is not invertible if we have no information about the exact molecular continuum states. However, if we have some knowledge of the molecule \emph{a priori}, an approximate form of continuum state $\psi_{\mathbf{k}}^{app}(\mathbf{r})$ including the influence of the molecular potential can be assumed. It is reasonable to do so because in order to retrieve the bound state from the measured dipole matrix element, one should make an initial guess of the continuum state. After obtaining the appropriate continuum state $\psi_{\mathbf{k}}^{app}(\mathbf{r})$, Eq. (7) becomes
\begin{eqnarray}
\mathbf{d}(\mathbf{k})&=&(2\pi)^{-3/2} \int d^{3}\mathbf{k'}\varphi_{\mathbf{k}}^{app}(\mathbf{k'})\nonumber\\&&\times[\int d^{3}\mathbf{r}  \psi_{0}(\mathbf{r})\mathbf{r}\exp(i\mathbf{k'}\cdot\mathbf{r})],
\end{eqnarray}
where
\begin{eqnarray}
\varphi_{\mathbf{k}}^{app}(\mathbf{k'})=(2\pi)^{-3/2}\int d^{3}\mathbf{r} \psi_{\mathbf{k}}^{app} (\mathbf{r}) \exp(-i\mathbf{k'}\cdot\mathbf{r}).
\end{eqnarray}
We can define the matrix of Fourier-space mapping as
\begin{eqnarray}
\mathbb{S}=\left(
\begin{array}{cccc}
 \varphi_{\mathbf{k_1}}^{app}(\mathbf{k_1'}) &\varphi_{\mathbf{k_2}}^{app}(\mathbf{k_1'})  & ...&\varphi_{\mathbf{k_n}}^{app}(\mathbf{k_1'})  \\
                                                                                                                                      \\
 \varphi_{\mathbf{k_1}}^{app}(\mathbf{k_2'}) &\varphi_{\mathbf{k_2}}^{app}(\mathbf{k_2'})  & ...&\varphi_{\mathbf{k_n}}^{app}(\mathbf{k_2'})   \\
 .   & .                                         & .& .   \\
 .   & .                                         & .& .   \\
 .   & .                                         & .& .  \\
 \varphi_{\mathbf{k_1}}^{app}(\mathbf{k_n'}) &\varphi_{\mathbf{k_2}}^{app}(\mathbf{k_n'})  & ...&\varphi_{\mathbf{k_n}}^{app}(\mathbf{k_n'})
\end{array}
\right),
\end{eqnarray}
where $\varphi_{\mathbf{k}}^{app}(\mathbf{k'})\sim\langle\exp(i\mathbf{k'}\cdot\mathbf{r})|\psi_{\mathbf{k}}^{app}(\mathbf{r})\rangle$ is the projection coefficient. The sketch of the obtaining of the Fourier-space mapping matrix can be seen in Fig. 2. For each component $\psi_{\mathbf{k}}^{app}(\mathbf{r})$ of the continuum wave function with the specific momentum $\mathbf{k}$, it can be projected onto a set of complete bases composed of plane waves and form a column vector. This column vector can be interpreted as the momentum-space representation of $\psi_{\mathbf{k}}^{app}(\mathbf{r})$. Putting the column vectors of all $\varphi_{\mathbf{k}}^{app}(\mathbf{k'})$ together, we can obtain the projection coefficient matrix, \emph{i.e.}, the transformation matrix $\mathbb{S}$. Thus Eq. (7) can be rewritten as
\begin{eqnarray}
\mathbf{d}(\mathbf{k})=(2\pi)^{-3/2}\mathbb{S} [\int d^{3}\mathbf{r}  \psi_{0}(\mathbf{r})\mathbf{r}\exp(i\mathbf{k'}\cdot\mathbf{r})].
\end{eqnarray}
Thereby, the molecular orbital can be reconstructed based on
\begin{eqnarray}
\psi_{0}(\mathbf{r})=\frac{\mathfrak{F}_{k\rightarrow r}[\mathbb{S}^{-1}\mathbf{d}(\mathbf{k})]}{\mathbf{r}}.
\end{eqnarray}
\begin{figure}
\centerline{
\includegraphics[width=8cm]{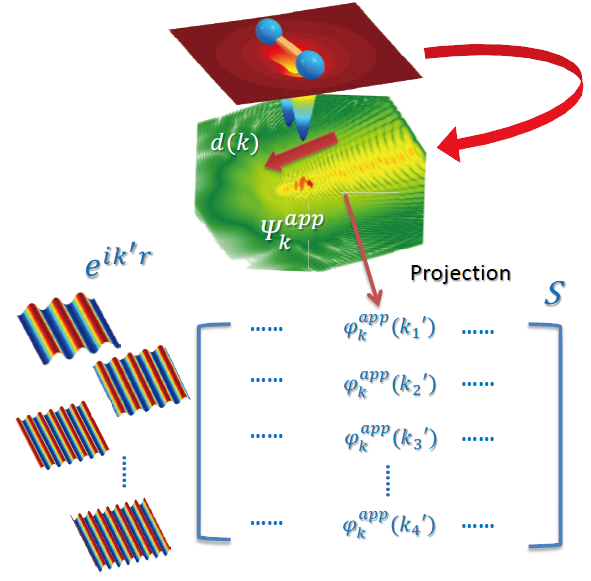}}
\caption{The sketch of the calculation of transformation matrix $\mathbb{S}$. The continuum wave function $\psi_{\mathbf{k}}^{app}(\mathbf{r})$ with the momentum $\mathbf{k}$ is projected onto a set of complete bases composed of plane waves. The projection coefficients $\varphi_{\mathbf{k}}^{app}(\mathbf{k'})$ can be calculated according to $\varphi_{\mathbf{k}}^{app}(\mathbf{k'})=\langle\exp(-i\mathbf{k}\cdot\mathbf{r})|\psi_{\mathbf{k}}^{app}(\mathbf{k'})\rangle$.  Putting all the projection coefficients together, the transformation matrix $\mathbb{S}$ can be obtained.}
\end{figure}

To summarize, with the amplitude of the continuum electron wave packet $a_{ewp}(\omega)$ factored out, the transition dipole matrix element $\mathbf{d}(\mathbf{k})$ is obtained. Then by projecting the continuum state $\psi_{\mathbf{k}}^{app}(\mathbf{r})$ onto plane waves, the transformation matrix $\mathbb{S}$ which defines the Fourier-space mapping from the desired quantity $\int d^{3}\mathbf{r} \psi_{0}(\mathbf{r})\mathbf{r}\exp(i\mathbf{k}\cdot\mathbf{r})$ to $\mathbf{d}(\mathbf{k})$ can be calculated. Finally, by inversing the mapping and performing the inverse Fourier transform, the molecular orbital can be successfully reconstructed beyond the plane wave approximation.

In the following, we will demonstrate this tomographic theory to reconstruct the symmetric $3\sigma_{g}$ HOMO of $\mathrm{N_2}$ molecule. We use a ten-cycle linearly polarized laser pulse with ¡°flat-top¡± envelope with three cycles rising and falling linearly and four cycles keeping constant. To minimize the multielectron effects which emerges from dynamical interference between harmonics generated from different molecular orbitals, we apply a mid-infrared laser pulse with a wavelength of $1200$ $\textrm{nm}$ and low intensity of $1.0\times 10^{14}\mathrm{W/cm^2}$ \cite{Vozzi2}. The exact $3\sigma_{g}$ HOMO of $\textrm{N}_2$ is calculated with Gaussian 03 \emph{ab initio} code \cite{Gaussian} and the two-dimensional projection of this orbital is shown in Fig. 3(a). We calculated the HHG spectra using a frequency-domain model \cite{Worner}, similar to the QRS theory \cite{Lin2,Jin}. The induced dipole moment can be expressed as
\begin{eqnarray}
D(\omega,\theta)=a_{ion}(\omega,\theta)a_{ewp}(\omega)d(\omega,\theta),
\end{eqnarray}
where $\theta$ is the angle between the laser-field polarization and the molecular axis. The factor $a_{ion}(\omega,\theta)$ represents the angular variation of the strong-field ionization rate calculated by MO-ADK theory \cite{Tong}, $a_{ewp}(\omega)$ describes the propagation amplitude of the re-colliding electron wave packet and $d(\omega,\theta)$ is the transition matrix element which is obtained from \emph{ab initio} quantum scattering calculations using EPOLYSCAT \cite{Lucchese1,Lucchese2}. In our calculation, HHG data are obtained between $0^{\circ}$ and $90^{\circ}$ with angular step of $\Delta\theta=5^{\circ}$. Spectra at the remaining angles are complemented exploiting the prior symmetry knowledge of the HOMO. The spectral range used in the tomography procedure is from $20$ eV to $90$ eV with a step of $2$ eV.

Because $a_{ewp}(\omega)$ only depends on the driving laser field, not on the structure of the target, it can be calibrated using a reference atom $\mathrm{Ar}$ with the same ionization potential of $\mathrm{N_2}$. The transition dipole $d(\omega,\theta)$ is given by
\begin{eqnarray}
d_{x/y}(\omega,\theta)&=&\frac{1}{\eta(\theta)}\frac{A^{mol}_{x/y}(\omega,\theta)}{A^{ref}(\omega)}d^{ref}(\omega)\nonumber\\&&\times\exp[i\phi^{mol}_{x/y}(\omega,\theta)-i\phi^{ref}(\omega)].
\end{eqnarray}
Here, $A^{mol}$, $A^{ref}$, $\phi^{mol}$ and $\phi^{ref}$ denote the amplitude and phase of the harmonics generated from molecules and reference atoms, respectively. $(x, y)$ are the coordinates of the molecular reference
frame with the internuclear axis along x. $\eta$ is a scaling factor representing the $\theta$-dependence of square root of the ionization probability \cite{Haessler2}.

Here we use the two-center Coulomb waves with outgoing boundary conditions as the molecular continuum wave function \cite{Cippina}. This wave function is the solution of the two-body Coulomb continuum problem which takes into account the main Coulomb effects on the re-colliding wave packet. The TCC can be written as
\begin{eqnarray}
\psi_{\mathbf{k}}^{TCC}(\mathbf{r})=(2\pi)^{-3/2}\exp(i\mathbf{k}\cdot\mathbf{r})\textit{M}(\mathbf{k},\mathbf{r_1})\textit{M}(\mathbf{k},\mathbf{r_2}),
\end{eqnarray}
with
\begin{eqnarray}
\textit{M}(\mathbf{k},\mathbf{r})=\exp(\frac{\pi\nu}{2})\Gamma(1-i\nu)_1F_1[i\nu,1,i(k r-\mathbf{k}\cdot\mathbf{r})].
\end{eqnarray}
Here, $\mathbf{r_1}=\mathbf{r}+\mathbf{R}/2$ and $\mathbf{r_2}=\mathbf{r}-\mathbf{R}/2$. $\mathbf{R}$ is internuclear distance and $\nu=Z/k$ is the Sommerfeld parameter where Z is the effective ion charge. We set $Z=0.5$ for each ion to match the condition that the $\mathrm{N_2^+}$ ion acts on the recolliding electron with the effective charge of $+1$ asymptotically. Inserting Eq. (15) into Eq. (9), we obtain the momentum-space representation of TCC and thus the transformation matrix $\mathbb{S}$. This form of TCC usually appears in collision physics and can be calculated using the Norsdieck method \cite{Yudin,Nordsieck}.

\begin{figure}
\centerline{
\includegraphics[width=10cm]{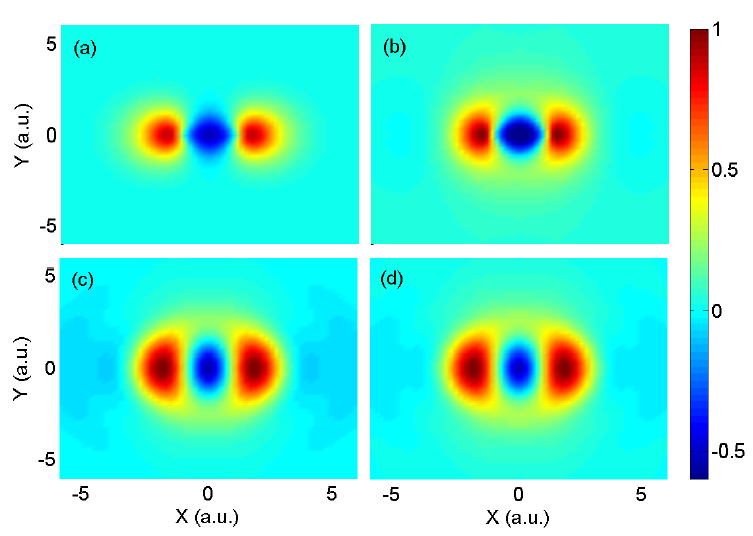}}
\caption{Two-dimensional projection of the $3\sigma_{g}$ HOMO of $\mathrm{N_2}$ (a) calculated by Gaussian 03 \emph{ab initio} code, (b) reconstructed using the proposed method, (c) reconstructed using the plane wave approximation and (d) reconstructed using the method in Ref.\cite{Vozzi2}. }
\end{figure}

Substituting the dipole $d(\omega,\theta)$ and matrix $\mathbb{S}$ into Eq. (12), the HOMO of $\mathrm{N_2}$ is reconstructed as shown in Fig. 3(b) together with the \emph{ab initio} orbital in Fig. 3(a). Using the same input for HHG spectra, we also present the molecular orbitals reconstructed based on PWA and the method in Ref.\cite{Vozzi2}, as shown in Fig. 3(c) and (d). All the three reconstructed HOMO images reproduce the main features of the target molecule which show alternating positive and negative lobes and two nodal planes along the y direction. The distance between the two nitrogen ions, estimated as the distance between the nodes of the HOMO lobes along the molecular axis, is about $1.02$ $\mathrm{a.u.}$, in good agreement with the \emph{ab initio} result. From Fig. 3, one can clearly see that the quality of the retrieved orbital using our method is much better than those two methods, it recovers the fine structures of the target molecule whereas additional structures appear in Fig. 3(c) and (d) which do not exist in the exact HOMO image.

The quality of the retrieved orbital using our method is further verified from the slices of the orbitals shown in Fig. 4. The slice using our method well matches the exact orbital, the value of the wavefunction almost coincides with the exact one, while the other two slices using PWA and method in Ref.\cite{Vozzi2} show obvious deviations from the exact orbital. Our retrieved HOMO image effectively removes the artificial term caused by the molecular potential and provides a more reliable reproduction of the target orbital. Therefore our proposed method has a significant improvement over previous methods, especially for sophisticated molecules since the additional structures may bring misleading information of the recovered image of the complicated molecular orbital. In addition, the tomographical images can be further improved if one adopts continuum wave function calculated by more elaborate theories such as the independent atomic center approximation \cite{Hasegawa} and single-scattering theories \cite{Kera}.
\begin{figure}
\centerline{
\includegraphics[width=8cm]{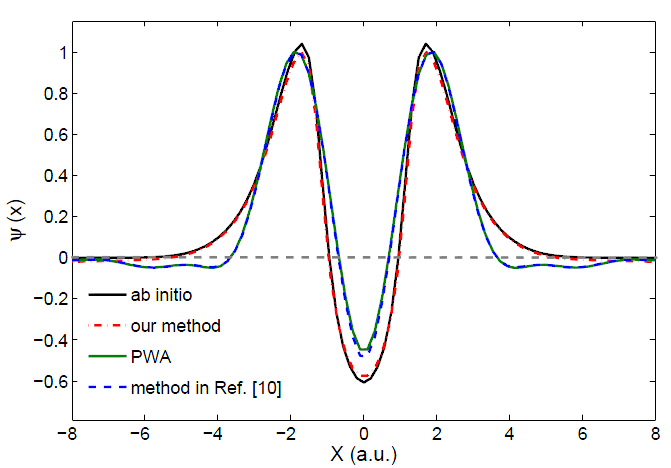}}
\caption{Slices obtained by cutting along the internuclear axis for the \emph{ab initio} (black solid line) and reconstructed orbitals using the proposed method (red dashed-dotted line), the plane wave approximation (green solid line) and the method in Ref.\cite{Vozzi2} (blue dashed line).   }
\end{figure}

It is worth noting that before our theory is proposed, the use of PWA is essential in the original MOT procedure. However, owing to the fact that PWA ignores the essential properties of the continuum wave function such as the distortion by the Coulomb potential, it brings a long-standing controversy about the validity of MOT since it was proposed. Our theory breaks through the restriction of PWA, and allows retrieving the molecular orbital directly using continuum wave function. Within the proposed theory, the continuum wave function given in any forms can be used and the accuracy of the tomographical images is no longer limited by the using of PWA. In this case, the queries on the theoretical basis of MOT which result from the use of PWA is solved.

In summary, we have developed a theory of molecular orbital tomography directly using continuum wave functions. In contrast to the commonly used plane wave approximation, our treatment accounts for the modulation of the continuum wave function caused by the molecular potential. Within our approach, the continuum wave function can be decoded using the momentum-space representation of this wave function, and the reversibility of the mapping relationship from the molecular orbital to the high-order harmonic spectra is maintained. According to our theory, any forms of continuum wave function can be used in the retrieval procedure. As a demonstration, we have reconstructed the $3\sigma_{g}$ HOMO of $\mathrm{N_2}$ molecule using two-center Coulomb waves (TCC) as the continuum wave function. With this approach, quantitative agreement between the reconstructed result and the \emph{ab initio} one can be reached. Our theory clarifies the long-standing controversy about the validity of MOT and strengthen the theoretical basis of MOT. Our formulation could be more useful to retrieve orbitals of sophisticated molecules, because the continuum state is very sensitive to the molecular potential and thus the plane waves are not adequate to describe it for sophisticated molecules.

We gratefully acknowledge R. R. Lucchese for providing us EPLOYSCAT and C. D. Lin, Z. Chen and  C. Jin for helpful discussions. This work was supported by the NNSF of China under Grants No. 11234004 and 61275126, and the 973 Program of China under Grant No. 2011CB808103.

\end{document}